\documentclass[twocolumn,showpacs,preprintnumbers,prd]{revtex4}
\usepackage{amsmath,amssymb}

\usepackage[dvips]{graphicx}
\usepackage{epstopdf}

\usepackage{dcolumn}
\usepackage{bm}

\newcommand{\ua}{$\mathrm{U}_{\rm A}(1)$ }

\newcommand{\rmi}{\mathrm{i}}
\newcommand{\rmd}{\mathrm{d}}
\newcommand{\rme}{\mathrm{e}}
\newcommand{\gs}{g_{\mathrm{S}}}
\newcommand{\gd}{g_{\mathrm{D}}}

\begin{document}

\preprint{}
\title{Model analysis of thermal UV-cutoff effects
 on the chiral critical surface \\at finite temperature and chemical potential}

\author{Jiunn-Wei Chen}
\email{jwc@phys.ntu.edu.tw}
\affiliation{Department of Physics
 and Center for Theoretical Sciences,
 National Taiwan University, Taipei 10617, Taiwan}%

\author{Hiroaki Kohyama}
\email{kohyama@phys.sinica.edu.tw}
\affiliation{Department of Physics,
Chung-Yuan Christian University, Chung-Li 32023, Taiwan and\\
Institute of Physics, Academia Sinica, Taipei 115, Taiwan and\\
Physics Division, National Center for Theoretical Sciences, Hsinchu 300, Taiwan
}

\author{Udit Raha}
\email{udit.raha@unibas.ch}
\altaffiliation{Present affiliation: Department of Physics and Astronomy,
University of South Carolina, Columbia, SC 29208, USA }
\affiliation{Department of Physics
and Center for Theoretical Sciences,
National Taiwan University, Taipei 10617, Taiwan
}


\begin{abstract}
We study the effects of temporal UV-cutoff on the chiral critical
surface in hot and dense QCD using a chiral effective model.
Recent lattice QCD simulations indicate that the curvature of the
critical surface might change toward the direction in which the first
order phase transition becomes stronger on increasing the
number of lattice sites. To investigate this effect on the critical
surface in an effective model approach, we use the Nambu-Jona-Lasinio
model with finite Matsubara frequency summation. We find that
qualitative feature of the critical surface does not alter appreciably
as we decrease the summation number, which is unlike the case what is 
observed in the recent lattice QCD studies. This may either suggest 
the dependence of chemical potential on the coupling strength or due to 
some additional interacting terms such as vector interactions which 
could play an important role at finite density.
\end{abstract}

\pacs{12.38.Aw,11.10.Wx,11.30.Rd,12.38.Gc}
\maketitle

\section{Introduction}~~
The study of the chiral critical point (CP) in the phase diagram of 
hot and dense quark matter is one of the central issues in Quantum 
Choromodynamics (QCD)~\cite{Fukushima:2010bq}.
While, it is widely accepted that the QCD  
phase transitions concerning chiral symmetry restoration and color  
deconfinement are crossovers with increasing temperature $T$ for  
small chemical potential $\mu \simeq 0$, the order of the phase  
transitions along the $\mu$ direction for small T is still under  
considerable speculation. Model analysis indicate that a first-order phase
transition occurs with increasing $\mu$ and for small $T$~\cite{Asakawa:1989bq}.
The above observations lead us to expect the existence of a critical
point located at the end of the first order line in the phase diagram
for some intermediate values, $T_E$ and $\mu_E$. 

A first principle determination of the phase diagram by solving QCD itself is
difficult due to the strongly interacting nature of matter at
low-energies/temperatures which significantly restricts the range of
applicability of perturbative calculations. We must then rely on 
non-perturbative techniques such as lattice QCD (LQCD) or some low-energy 
effective theories of QCD. The LQCD simulations are known to be
a viable approach for microscopic calculations in QCD, and have recently 
reached a reliable level at finite $T$ and $\mu = 0$~\cite{Philipsen:2008gf}.
However, these simulations are not yet able to provide a conclusive
understanding of the QCD phase diagram due to severe limitations posed by
the well-known ``sign-problem'' at finite $\mu$, and difficulties dealing
with small quark masses, though only very approximate  methods are available
for simulations at small $\mu$ values~\cite{deForcrand:2010ys}. Thus, it is
important to develop low-energy effective  models that may show consistency
with lattice results and can be extrapolated into regions not accessible
through simulations. Among them, the local Nambu-Jona-Lasinio (NJL)
model~\cite{NJL} and its proposed extended version to include coupling
of quarks to Polyakov-loops, the so-called Polyakov-loop NJL (PNJL)  
model~\cite{Fukushima:2003fw}, are useful to study the quark system at finite 
$T$ and/or $\mu$. Such effective models share the same symmetry properties 
of QCD and successfully describe the observed meson properties and chiral
dynamics at low-energies (see, e.g.,~\cite{Hatsuda:1994pi,Klevansky:1992qe,%
Buballa:2005rept,Fukushima:2008wg}).

Using the framework of (P)NJL model we search the CP and analyze the order
of the chiral phase transition by varying the current quark masses
($m_u$, $m_d$, $m_s$). We usually set, for simplicity, $m_d=m_u$
and investigate the phase transitions in the $m_{u}$-$m_s$-$\mu$
space. Renormalization Group (RG) analysis of chiral models at $\mu\simeq0$
conclude that there is no stable infra-red (IR) fixed-point for quark flavors 
$N_F>\sqrt{3}$~\cite{Pisarski:1983ms}, indicating that the thermal phase
transition is of fluctuation induced first order for two or more flavors
realized in the chiral limit $m_{u,d,s}=0$. However, the transition becomes 
a crossover for intermediate quark masses because of explicitly broken chiral 
and center symmetries. Hence, it is naturally expected that there should be a 
``critical boundary'' separating the regions of the first order phase 
transition and crossover between small and intermediate quark masses. 
Although LQCD results support the above model picture qualitatively, there
had been a huge quantitative difference between the two kinds of analyses,
even at zero chemical potential where the LQCD does not suffer from the 
sign-problem; the value of the critical mass obtained in the (P)NJL model is
about one order of magnitude smaller than the value in the LQCD analyses.
Inspired by recent works reporting that the critical mass may become
smaller when the number of lattice sites is increased~\cite{deForcrand:2007rq,%
Philipsen:2009dn},  the present authors studied the critical boundary in the
(P)NJL model with finite Matsubara frequency summation $N$ at zero chemical
potential~\cite{Chen:2009mv}. There it was
found that the critical mass actually becomes larger if one decreases the
Matsubara summation number $N$ (see, Eq.(\ref{trick})), thereby showing the
correct tendency to explain the quantitative difference between the LQCD and
(P)NJL model results.

In this Letter, our goal is to study the critical boundary for all values of
the chemical potential, which may eventually tell us the location of the CP
in the QCD phase diagram. As already mentioned that since an {\it ab initio}
determination of the critical point in QCD is a distant hope, nevertheless,
it is worth studying the (P)NJL model with finite frequency  summation at
finite chemical potential that should capture the essential qualitative
features of the results expected in lattice studies. Note that in the
$m_{u}$-$m_s$-$\mu$ space, the critical boundary becomes a surface called the
``critical surface''. More precisely, through this study we would like to
investigate the qualitative behavior of this critical surface whose shape can
critically determine whether the CP exists. We can then compare our results
with the recent lattice predictions.

\section{Model set up}~~
The NJL model Lagrangian in the 3 flavor system is written by
\begin{align}
 \mathcal{L}_{\mathrm{NJL}} &= \bar{q}\left( \rmi\partial\!\!\!/
  - \hat{m}\right)q + \mathcal{L}_4 + \mathcal{L}_6 ,
\label{LNJL} \\
 \mathcal{L}_4 &= \frac{\gs}{2} \sum_{a=0}^8 \left[
  \left( \bar{q}\lambda_a q\right)^2
  + \left( \bar{q}\,\mathrm{i}\gamma_5 \lambda_a q \right)^2
  \right] ,
\label{L_4} \\
 \mathcal{L}_6 &= -\gd \left[ \det\bar{q}_i (1-\gamma_5) q_j 
  + \text{h.c.\ } \right] .
\label{L_6}
\end{align}
Here $\hat{m}$ is the diagonal mass matrix $(m_u,\,m_d,\,m_s)$ in the flavor
space which explicitly breaks the chiral symmetry. $\mathcal{L}_4$ is the
4-fermion contact interacting term with coupling constant $\gs$, and $\lambda_a$
is the Gell-Mann matrix in the flavor space with
$\lambda_0=\sqrt{2/3}\,{\rm diag}(1,\,1,\,1)$. $\mathcal{L}_6$ is a 6-fermion
interaction term called the  Kobayashi-Maskawa-t'Hooft interaction whose
coupling strength is  $\gd$~\cite{Kobayashi:1970ji}. The subscripts ($i,\,j$)
indicate the  flavor indices and the determinant runs over the flavor space. 
This term is introduced to explicitly break the \ua symmetry.

To study the chiral dynamics, we solve the gap equations which
are derived through differentiating the thermodynamic potential
$\Omega$ by the order parameters of the model:
\begin{equation}
 \frac{\partial\, \Omega}{\partial\, m_i^*}=0 \quad ; \quad i=u,d,s
\label{gap0}
\end{equation}
The order parameters $m_i^*$  are the constituent quark masses.
Since, we set $m_d=m_u$ in our analysis, this should lead to the 
isospin symmetric result $m_d^*=m_u^*$, reducing the number of gap equations
from three to two. The thermodynamic potential $\Omega$ is defined by
$\Omega \equiv -\ln Z/(\beta V)$, where $Z$ is the partition
function, $\beta(\equiv 1/T)$ is the inverse temperature and $V$ is the
volume of the thermal system.

In the mean-field approximation, after some algebra, we arrive at the
following expressions for the gap equation:
\begin{equation}
 \begin{split}
  m_u^* &= m_u + 2\rmi \,\gs\, N_c \mathrm{tr}S^u
   - 2\gd N_c^2 (\mathrm{tr}S^u)(\mathrm{tr}S^s) , \\
  m_s^* &= m_s + 2\rmi \,\gs\, N_c \mathrm{tr}S^s
   - 2\gd N_c^2 (\mathrm{tr}S^u)^2 ,
\end{split}
\label{gap}
\end{equation}
where $N_c(=3)$ is the number of colors and tr$S^i$ is the chiral
condensate written explicitly as
\begin{equation}
 \rmi\; \mathrm{tr}S^i = 4m_i^* \int \! \frac{d^3p}{(2\pi)^3}\;
  (\rmi\,T)\!\! \sum_{n=-\infty}^{\infty}
  \frac{\rmi}{(\rmi\,\omega_n)^2 - E_i^2} \, .
\label{trace}
\end{equation}
Here $w_n=\pi T (2n+1)$ is the Matsubara frequency and
$E_i=\sqrt{{\bf p}^2+m_i^{*\,2}}$ is the energy of the quasi-particle.
A detailed calculation for deriving the gap equations Eq.(\ref{gap}) is
clearly presented in the review paper~\cite{Klevansky:1992qe}.

To study the UV-cutoff effects in the model, we cut the higher
frequency modes in the Matsubara sum as employed in~\cite{Chen:2009mv},
\begin{equation}
\label{trick}
 \sum_{n=-\infty}^\infty \longrightarrow \sum_{n=-N}^{N-1} .
\end{equation}
This model has five free parameters $\{m_{u},\,m_s,\,\Lambda,\,\gs,\,\gd\}$:
two current quark masses, a 3-dimensional momentum cutoff, a four-fermion and
a six-fermion coupling constants. Following~\cite{Hatsuda:1994pi}, we set
$m_{u}=5.5$MeV fixed for all values of $N$, while the remaining four
parameters are fitted each time by using the following physical observables
\begin{align}
 &m_{\pi}=138 \text{ MeV}, && f_{\pi}=93 \text{ MeV}, \nonumber\\
 &m_K=495.7 \text{ MeV}, && m_{\eta^{\prime}}=958 \text{ MeV}. \nonumber
\end{align}
The parameter fitting for various $N$ has been done in~\cite{Chen:2009mv},
and we employ the same values in this analysis which are again displayed 
in Tab.~\ref{parameters} for the convenience of the reader.


\begin{table}[tbp]
 \begin{tabular}{c@{\hspace{0.5cm}}c@{\hspace{0.5cm}}c%
@{\hspace{0.5cm}}c@{\hspace{0.5cm}}c}
 \hline\hline
$N$ & $m_s$ (MeV) & $\Lambda$ (MeV) & $\gs\Lambda^2$ & $\gd\Lambda^5$ \\
 \hline
$15$ & $134.7$ & $631.4$ & $4.16$ & $12.51$ \\ 
$20$ & $135.0$ & $631.4$ & $4.02$ & $11.56$ \\ 
$50$ & $135.3$ & $631.4$ & $3.82$ & $10.14$ \\ 
$100$ & $135.4$ & $631.4$ & $3.75$ & $9.69$ \\ 
c$\infty$ & $135.7$ & $631.4$ & $3.67$ & $9.29$ \\
 \hline\hline
\end{tabular}
\caption{The various fitted parameters for different $N$~\cite{Chen:2009mv}.}
\label{parameters}
\end{table}


\section{Chiral Critical Surface}~~
The main purpose of this Letter is to determine the chiral critical 
surface in the (P)NJL model with finite Matsubara summation.
The critical surface is the set of all
critical points in the $m_u$-$m_s$-$\mu$ space which are analyzed by scanning 
the space for discontinuities of the chiral condensate. It should be noted that
for each value of $N$, we treat both the current quark masses $m_u$ and $m_s$ 
as free parameters in obtaining the critical surface once the other
parameters, namely, $\Lambda$, $\gs$ and $\gd$ are determined by fitting to
the physical parameters, as shown in Tab.~\ref{parameters}. Of course, we will 
eventually be interested in the case of the real (physical) current quark masses
$m_{u}\simeq 5.5$MeV and $m_s\simeq 136$MeV, in order to determine the
possible existence/non-existence of the CP through our model analysis.
Because our motivation is to make a direct comparison of our results with that of
LQCD where the simulations are mainly performed in the $m_u=m_s$ symmetric
case at finite $\mu$, we shall also consider this case.
In the actual numerical calculations, we scouted out the critical masses
($m_{uc}$,$m_{sc}$) for each $\mu$ by searching for discontinuities in the
solutions of the gap equations Eq.(\ref{gap}) in the entire $m_u$-$m_s$-$\mu$ 
space.

The LQCD and model studies indicate a crossover realized at $\mu=0$ for
physical current quark masses. This means that the curvature of the critical 
surface will tell us whether the CP is favored in the phase diagram. To be more
concrete, if the region of the first order phase transition expands with
increasing $\mu$, the physical quark mass line will intersect with the critical 
surface and this will end up as a CP. If on the other hand, the first order
phase transition region shrinks with $\mu$, there is less chance of an
appearance of a the CP and a crossover transition will be favoured for the whole 
range of $T$ and $\mu$.

\begin{figure}[tbh]
\begin{center}
\includegraphics[width=1.03\columnwidth]{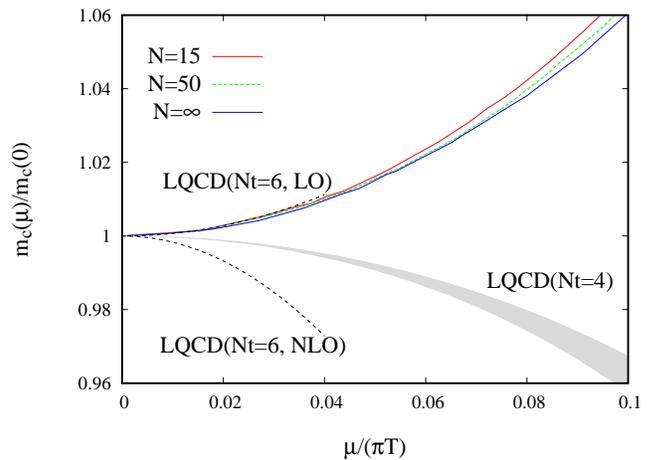}
\end{center}
\caption{$m_c(\mu)/m_c(0)$ vs $\mu/(\pi T)$ plot for different $N$ values
from the PNJL model, along with the corresponding results obtained from lattice 
simulations for $N_{t}=4$ and $N_{t}=6$.}
\label{slope_p}
\end{figure}

In the LQCD calculations, the curvature of the critical surface along the
$m_u=m_s$ symmetric line is analyzed by obtaining the critical mass $m_c$
through the following Taylor expansion formula
\begin{equation}
\begin{split}
  \frac{m_c(\mu)}{m_c(0)} = 1 + \sum_{k=1} c_k
    \left( \frac{\mu}{\pi T} \right)^{2k},
\end{split}
\label{Taylor}
\end{equation}
which so far yielded the following results: $c_1=-3.3(3)$, $c_2=-47(20)$
for $N_t=4$, and $c_1=7(14)$, $-17(18)$ (preliminary) for a leading order
(LO) and next-to-leading  order (NLO) extrapolation in $\mu^2$, respectively, for
$N_t=6$ where $N_t$ represents the number of the lattice sites in the temporal 
direction~\cite{Philipsen:2009dn}. These results are graphically represented in 
Fig.~\ref{slope} and \ref{slope_p} in the following sections which indicate that
the sign of the curvature has not yet been determined from lattice simulations.

\section{NJL model results}~~
\begin{figure}[tbh]
\hspace{-1.3cm}\includegraphics[width=1.15\columnwidth]{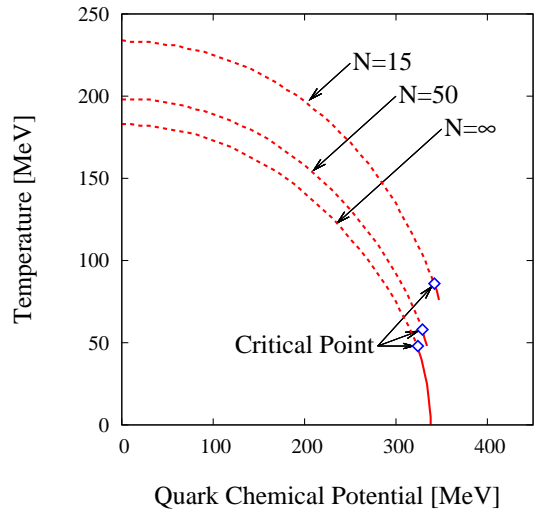}
\caption{Position of the CP in the $T$-$\mu$ phase diagrams from the NJL model 
with finite Matsubara summation for $N=15,\,50,\,\infty$. The dotted and the
solid lines represent crossovers and first order phase transitions, respectively. 
The $N=\infty$ case corresponds to the traditional NJL model.}
\label{pd}
\end{figure}
The phase diagrams of the NJL model for $N=15$, $50$ and $\infty$, 
respectively, are shown in Fig.~\ref{pd}. In fact, we have also studied the
model for other values of $N$, however, there were no significant qualitative 
differences and we prefer to select just the above three representative cases for graphical 
clarity. Here, it may be noteworthy mentioning that for values of $N<15$,
it becomes a matter of numerical challenge to perform simulations to determine
the phase boundaries.
So, henceforth, we shall be displaying our results only for the above three
cases. Note that the case $N=\infty$ corresponds to the traditional
NJL model. In drawing the phase diagrams, we apply the same 
criterion employed in~\cite{Fukushima:2008wg} where the phase transition or
crossover is defined by the condition, 
\begin{equation}
 \frac{\langle \bar{u}u \rangle }{\langle \bar{u}u \rangle_{T_0} }
 \biggl|_{T=T(\mu)}
=
 \frac{1}{2}\,,
\end{equation}
$\langle \bar{u}u \rangle_{T_0}$ being the expectation value of the chiral
condensate for the up quark at temperature $T_0$ and $\mu = 0$. We choose $T_0=0$ for the
$N=\infty$ traditional model, while we set $T_0=50$MeV for the case with
finite $N$ since the model is ill-defined for small $T$, as
discussed in~\cite{Chen:2009mv}. This is why we choose not to display the 
results for the small $T$ region where the curves are no longer physically reliable.

Here it is seen that the region below each of the curves that represents the 
chiral symmetry broken phase expands with decreasing $N$. This comes from the 
fact that the coupling constants become larger with decreasing $N$, being 
consistent with RG arguments that the coupling strength becomes
smaller when one considers the physics at higher momenta, i.e., for larger $N$.
When the coupling strength grows the chiral condensate tends to enlarge,
which can be easily seen from Eq.(\ref{gap}). Thus, it is naturally understood
why the transition temperature and chemical potential increase with a smaller
choice of the summation number $N$.

\begin{figure}[tbh]
\begin{center}
\includegraphics[width=1.03\columnwidth]{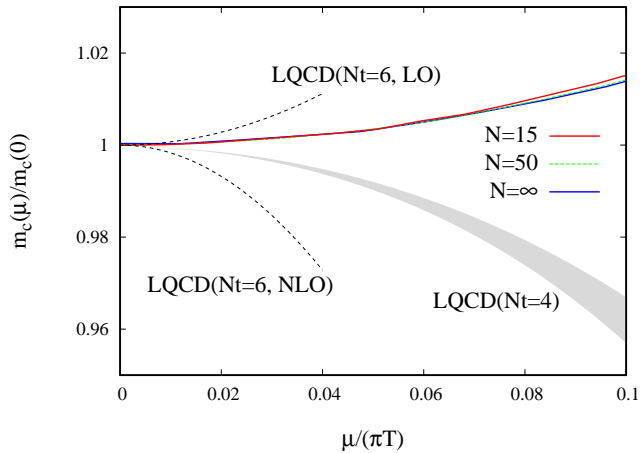}
\end{center}
\caption{$m_c(\mu)/m_c(0)$ vs $\mu/(\pi T)$ plot for different $N$ values
from the NJL model, along with the corresponding results obtained from lattice 
simulations for $N_{t}=4$ and $N_{t}=6$.}
\label{slope}
\end{figure}

In Fig.~\ref{slope}, we show the numerical results for $m_c(\mu)/m_c(0)$
as a function of $\mu/(\pi T)$, along with the corresponding results obtained
in the recent LQCD simulations for $N_{t}=4$ and $6$~\cite{Philipsen:2009dn}, respectively.
We see that
the slope of the curves tends to go up only very slightly on decreasing $N$ from
$N=\infty$ down to $N=15$.
On the other hand, the results from lattice simulations do not yield conclusive
results so far, as clearly revealed from the above figure, where the
sign of the curvature has not yet been constrained for the $N_t=6$
case. Thus, in comparison with the lattice studies, we find that the
qualitative behavior of our model results hardly changes by using
different $N$ values in our analysis. This means that there is indeed a
stark quantitative contrast between the NJL model results and the
lattice predictions. 

\begin{figure}[tbh]
\begin{center}
\includegraphics[width=1.0\columnwidth]{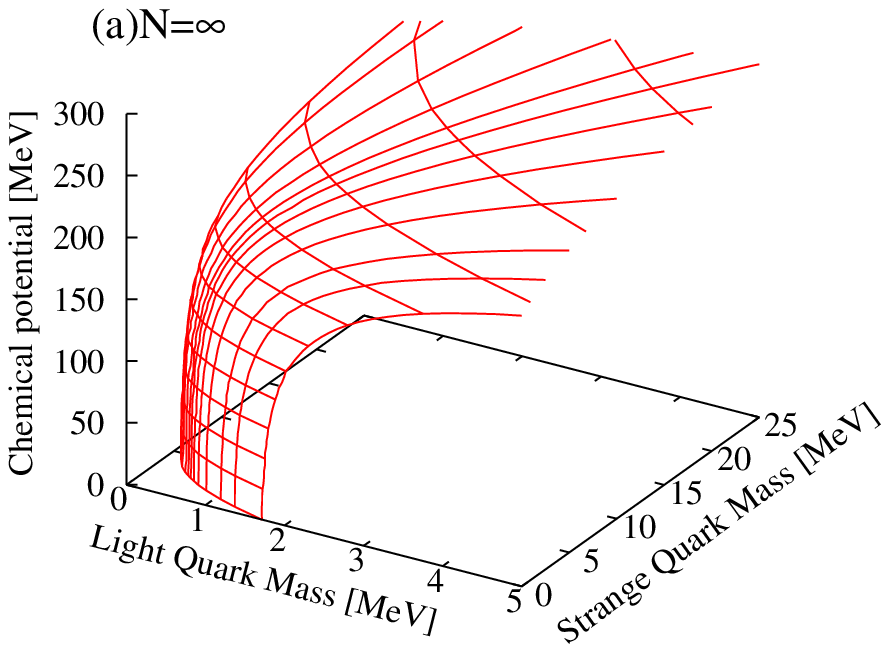}
\includegraphics[width=1.0\columnwidth]{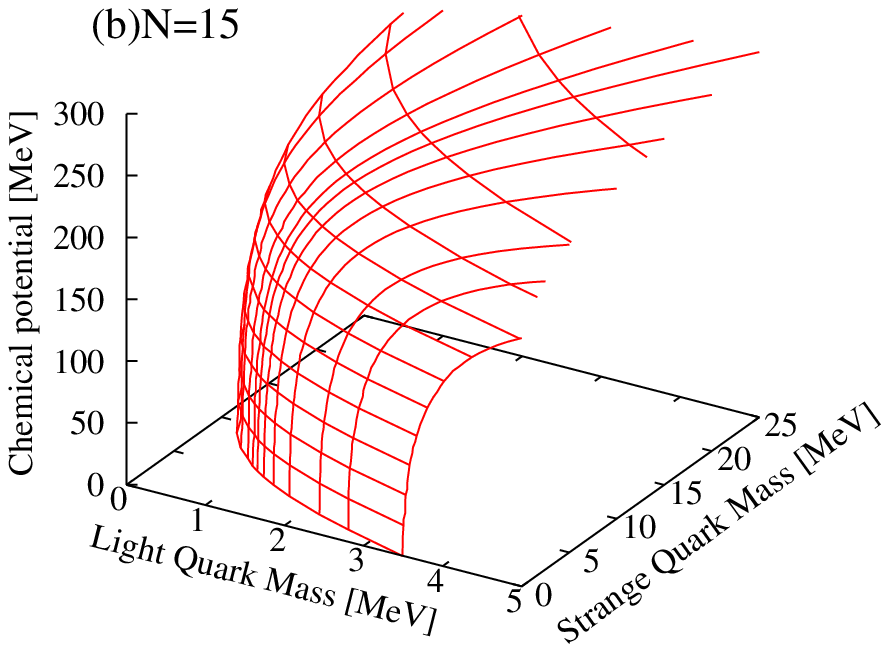}
\end{center}
\caption{The chiral critical phase surfaces for (a) $N=\infty$, and (b) $N=15$
from the NJL model in the $m_u$-$m_s$-$\mu$ space.}
\label{surface}
\end{figure}

As the final remark in this section, we display our results for the critical
surface obtained for the $N=\infty$ and $15$ cases in Fig.~\ref{surface}. It is
clearly seen that, although the values of the critical mass changes about factor
of two, the qualitative picture does not differ as we change the value of $N$.
Thus, the region of the first order phase transition expands with respect to $\mu$
for all $N$ values, since the effect of variation of $N$ is rather too nominal to
change the sign of the curvature of the critical surface. This would then mean that
the NJL model will always favour the CP with physical current quark masses at
some finite value of $\mu$, e.g., we obtain the CP when $\mu$ goes up 
around $\mu_c=324\,(342)\,$MeV for $N=\infty\,(15)$,
as exhibited in the phase diagram Fig.~\ref{pd}.

\section{The PNJL model extension}~~
It is also intriguing to study the critical surface in the
PNJL model, because of the closer resemblance to QCD which treats 
the chiral and deconfinement phase transitions simultaneously. 
In the PNJL model, the order parameter for the deconfinement phase 
transition is the Polyakov-loop and it is described by a global mean-field being
similar to the chiral condensation in the traditional NJL model.
The Lagrangian of the PNJL model is written
by~\cite{Fukushima:2008wg},
\begin{align}
 & \mathcal{L}_{\mathrm{PNJL}} = \mathcal{L}_0+\mathcal{L}_4
  +\mathcal{L}_6 + \mathcal{U}(\Phi,\Phi^*,T) \, , \\
 & \mathcal{L}_0 = \bar{q}\left(\rmi \partial \!\!\!/
  - \rmi\gamma_4 A_4 - \hat{m}\right)q \, , \\
 & \mathcal{U}(\Phi,\Phi^*,T) = -bT\bigl\{ 54\,\rme^{-a/T}\Phi\Phi^*
  \notag \\
 &\qquad +\ln\bigl[1 - 6\Phi\Phi^* + 4(\Phi^3+\Phi^{*\,3})
  - 3(\Phi\Phi^*)^2 \bigr] \bigr\}\, ,
\label{Polyakov_pot}
\end{align}
where $\mathcal{U}$ is the Polyakov-loop effective potential
and $\Phi$ and $\Phi^*$ are the traced Polyakov-
and the anti-Polyakov-loop, respectively. They are defined by
$\Phi=(1/N_c)\mathrm{tr}L$, $\Phi^*=(1/N_c)\mathrm{tr}L^\dagger$ with
$L=\mathcal{P} \exp[\rmi\int_0^\beta \!\rmd\tau A_4]$ and $A_4=\rmi A^0$.
There are several candidates for the Polyakov-loop potential in defining 
the PNJL model~\cite{Fukushima:2003fw,Fukushima:2008wg,Polyakov-loops},
and we adopt the strong-coupling inspired form of Eq.(\ref{Polyakov_pot})
following~\cite{Fukushima:2008wg}. In the above expressions, the parameter
$a$ solely parametrizes the strength of the Polyakov-loop condensate for
the deconfinement phase transition, while the parameter $b$ controls the  
relative strength of the mixing between the Polyakov-loop and chiral  
condensates, with a smaller value of $b$ signifying chiral phase transition
dominating over deconfinement. Here, the parameters $a$ and $b$ are set
as $a=664$ MeV, and $b\!\cdot\!\Lambda^{-3}=0.03$. Regarding the rest of
the model parameters, it is legitimate to use the same ones fixed in the
NJL model because the Polyakov-loop extension is likely to affect the system
only at finite temperatures comparable to the critical temperature $T_c$.
At much lower temperatures the chiral condensate is only very marginally
modified by the Polyakov loops.

\section{PNJL model results}~~
Let us now discuss the results in the PNJL model with finite frequency
summation.

\begin{figure}[tbh]
\hspace{-1.3cm}\includegraphics[width=1.15\columnwidth]{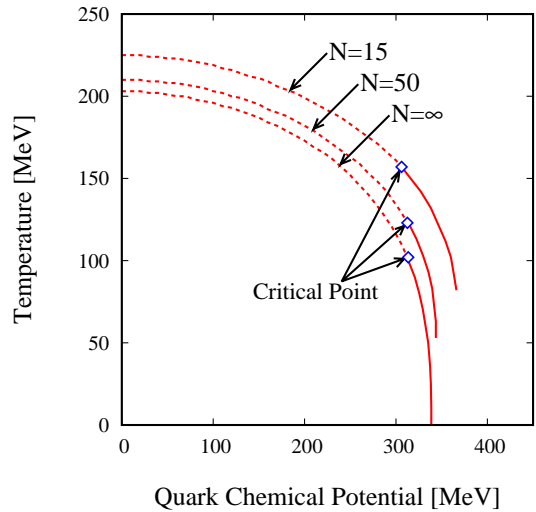}
\caption{Position of the CP in the $T$-$\mu$ phase diagrams from the PNJL
model  with finite Matsubara summation for $N=15,\,50,\,\infty$. The
dotted and the solid lines represent crossovers and first order phase
transitions, respectively. The $N=\infty$ case corresponds to the
traditional PNJL model.
}
\label{pd_p}
\end{figure}

In Fig.~\ref{pd_p}, we present the phase diagrams resulting from the PNJL model
with $N=15$, $50$ and $\infty$.
Here we see that the curves are very similar to the ones in Fig.~\ref{pd},
however, the critical temperatures are about a factor of two or more larger
than those obtained via the NJL model. This is almost the same quantitative difference
what is observed between the traditional ($N=\infty$) NJL and PNJL models.

We also display the corresponding curves for the ratio $m_c(\mu)/m_c(0)$ and
compare the results with the LQCD simulations. Again, the nature of the
results exhibit similar qualitative characteristics with the ones obtained in the NJL model;
the ratio does not change appreciably.
However, the ratios are numerically larger than that obtained with
the NJL case. This result can be interpreted as an effect of the Polyakov-loops
tending to suppress unphysical quark excitations below $T_c$~\cite{Fukushima:2008wg}.

\begin{figure}[tbh]
\begin{center}
\includegraphics[width=1.0\columnwidth]{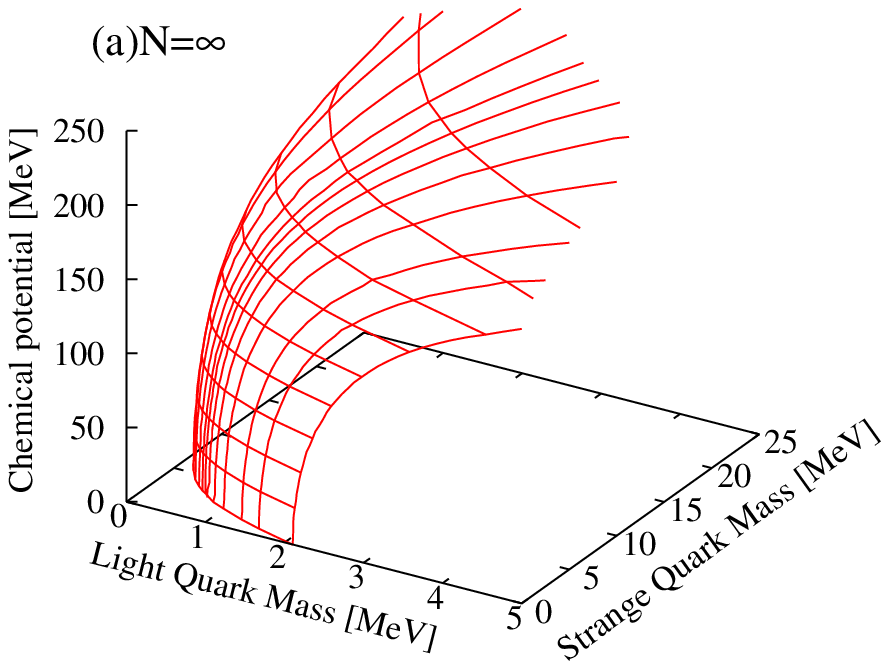}
\includegraphics[width=1.0\columnwidth]{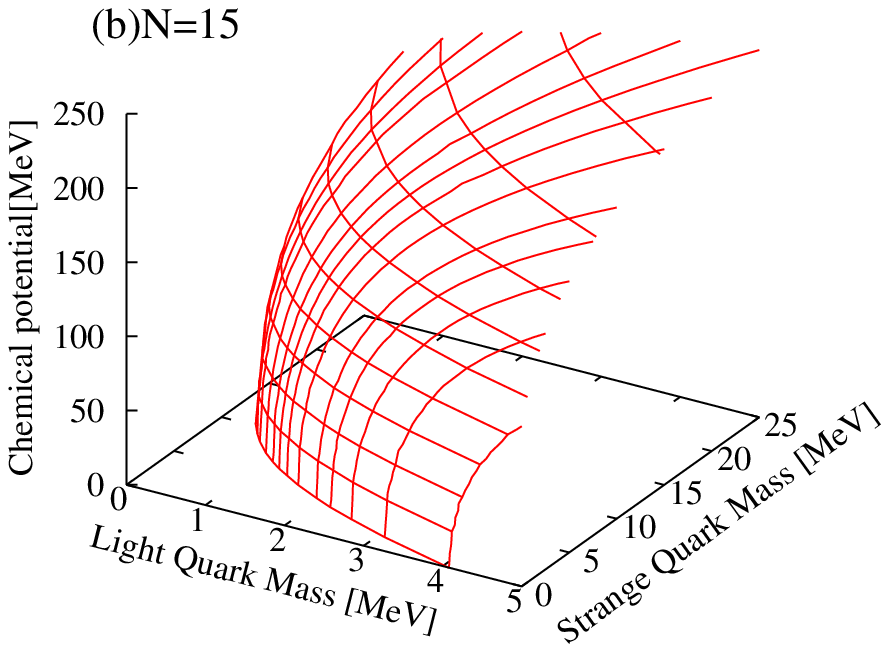}
\end{center}
\caption{The chiral critical phase surfaces for (a) $N=\infty$, and (b) $N=15$
from the PNJL model in the $m_u$-$m_s$-$\mu$ space.}
\label{surface_p}
\end{figure}

Finally, in Fig.~\ref{surface_p}, the critical surfaces in the PNJL model with $N=\infty$
and $15$ are displayed. We confirm almost the same qualitative
features in these two figures, quite similar to that previously found
in the NJL case. It is interesting to note that such qualitative similarity between
the NJL and PNJL model results is {\it a priori} non-trivial, since
the deconfinment phase transition order parameters, namely the Polyakov-loops $\Phi$ and
$\Phi^*$, respectively, may cause the system  with of chiral and deconfinment 
phase transitions to deviate significantly (both qualitatively and
quantitatively) from a system with only chiral phase transition, especially
in the vicinity of the CP.

\section{Summary and discussion}~~
In this Letter, we studied the UV-cutoff effects on the chiral critical surface
with two light and one heavy flavors using the (P)NJL model and found that 
its curvature is not appreciably affected by the finite Matsubara
frequency summation number. On the other hand, the current lattice simulations
are not decisive at the moment  which are beset with larger  
lattice cut-off effects than finite density effects, making continuum  
extrapolations doubtful. However, there is room for further precise  
lattice calculations in future with greater number of lattice sites  
and development of new techniques for finite-density simulations that  
may be necessary to make more definite conclusions.

As a final note, we point out the possibility of the temperature and
density dependence of the coupling constants that may become important at
high-energies, as we expect the couplings to run with respect to the energy scale.
However, in this Letter, we used constant values of the couplings $\gs$ and
$\gd$ which were fixed from fitting to physical quantities once and for all at
small temperature and zero chemical potential, to test the effects of the
temporal UV-cutoff. These dependencies may turn out to have crucial effect on
the critical surface when one considers the system at high densities, where
it is expected to be dominated by non-hadronic states. Thus, the location of
the CP is indeed
sensitive to the nature and magnitude of the coupling constants. In fact, it 
was actually found in~\cite{Chen:2009gv} that the $U_A(1)$ anomaly strength
modeled through the chemical potential dependent coupling $\gd$ may change 
the curvature of the critical surface, as well as its sign, resulting in a
characteristic ``back-bending'' of the critical surface as a function of
$\mu$. This reflects the fact that the density dependence of the coupling
strengths plays crucial role when investigating the chiral critical surface.

\begin{acknowledgments}
We are grateful to Ph. de Forcrand for suggesting us to perform this
calculation. HK likes to thank T. Inagaki and D. Kimura for fruitful
discussions. HK is supported by the grant NSC-99-2811-M-033-017 from
National Science Council (NSC) of Taiwan.
JWC and UR are supported by the NSC and NCTS of Taiwan. UR is also  
supported in part by the NSF grant of USA No. PHYS-0758114. He thanks  
the Institute of Mathematical Science Chennai and the Indian Institute  
of Technology Bombay, for their kind hospitality during the progress  
of this work. 
\end{acknowledgments}


\end{document}